\documentclass[10pt,prb,aps,superscriptaddress,twocolumn,longbibliography]{revtex4-1}
\usepackage{amssymb,amsmath}
\usepackage{graphicx}
\usepackage{color}
\usepackage{natbib}
\usepackage{hyperref}
\usepackage[english]{babel}
\usepackage[utf8]{inputenc}

\begin{document}

\newcommand\cyrtext[1]{{\fontencoding{T2A}\selectfont #1}}  
\newcommand{\df}[2]{\frac{\partial #1}{\partial #2}}             
\newcommand{\ds}[2]{\frac{{\partial}^2 #1}{\partial {#2}^2}}     
\newcommand{\tens}[1]{\hat{#1}}
\newcommand{\bas}[1]{\vec{e}_{#1}} 
\newcommand{\Gk}[2]{G_{\vec{k}}^{#1#2}} 
\newcommand{\refPR}[1]{[\onlinecite{#1}]}
\renewcommand{\cite}[1]{[\onlinecite{#1}]}
\newcommand{\red}[1]{{\color{red}#1}}
\newcommand{\av}[1]{\left< #1 \right>} 
\newcommand{\om}{\omega} 
\newcommand{\er}{\mathfrak{E}} 
\newcommand{\il}{\mathfrak{J}} 

\title{Nonlinear symmetry breaking in photo-metamaterials}
\author{Maxim~A.~Gorlach}
\email{m.gorlach@metalab.ifmo.ru}
\affiliation{ITMO University, Saint Petersburg 197101, Russia}

\author{Dmitry~A.~Dobrykh}
\affiliation{ITMO University, Saint Petersburg 197101, Russia}

\author{Alexey~P.~Slobozhanyuk}
\affiliation{ITMO University, Saint Petersburg 197101, Russia}

\author{Pavel~A.~Belov}
\affiliation{ITMO University, Saint Petersburg 197101, Russia}

\author{Mikhail Lapine}
\affiliation{School of Mathematical and Physical Sciences, University of Technology Sydney, NSW 2007, Australia}

\begin{abstract}

We design and analyze photo-metamaterials with each meta-atom containing both photodiode and light-emitting diode. Illumination of the photodiode by the light-emitting diode gives rise to an additional optical feedback within each unit cell, which strongly affects resonant properties and nonlinear response of the meta-atom. In particular, we demonstrate that symmetry breaking occurs upon a certain threshold magnitude of the incident wave intensity resulting in an abrupt emergence of second-harmonic generation, which was not originally available, as well as in the reduced third-harmonic signal.

\end{abstract}

\maketitle

\section{Introduction}\label{sec:Intro}

The flourishing field of nonlinear metamaterials~\refPR{Shadrivov-book,Lapine-2014,Lapine-2017} provides a wide variety of ways to implement artificial structures with unusual functionalities~\refPR{Zheludev} including magnetoelastic metamaterials~\refPR{Lapine-2011,Liu,Matsui}, bistable and self-tunable structures~\refPR{Zharov,Chen,Slob-AM}, metamaterials with nonlinear response tuned by the external static field~\refPR{Cai,Kang,Liu-AM-2016} and metamaterials exhibiting an interplay of electric and magnetic-type nonlinearities~\refPR{Rose2012,Rose-2013}.

One of the fascinating ideas in the field of nonlinear metamaterials is the implementation of additional interaction channel (or feedback) between the building blocks of artificial structure. For instance, it can be mechanical interactions~\refPR{Lapine-2011,Liu,Slob-AM} when the mechanical deformation of metamaterial affects the frequencies of electromagnetic resonances, or an additional optical channel realized by insertion of light sources and sensors into meta-atoms. In the case of latter structures further referred as photo-metamaterials the illumination of the sensor shifts the meta-atom resonance~\cite{Kapitanova-12,Slob2014}. As was demonstrated experimentally, the shift of the meta-atom resonance frequency due to optical feedback can be as large as $18$~MHz in microwave metamaterials~\refPR{Slob2014}.

We combine the idea of optical feedback with the well-celebrated concept of dynamic symmetry breaking 
manifesting itself in the lowering of the system symmetry under the applied external stimulus. Such a mechanism was studied in various physical contexts including Bose-Einstein condensates~\cite{Ostrovskaya,Aleiner}, topological photonics~\cite{Hadad}, nonlinear optical setups~\cite{Herring} and magnetoelastic metamaterials~\cite{Liu}.

    \begin{figure}[b]
    \begin{center}
    \includegraphics[width=0.95\linewidth]{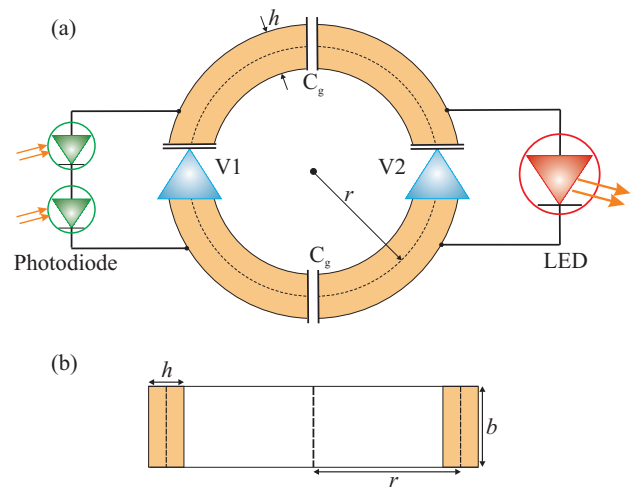}
    \caption{Electric scheme of the meta-atom with optical feedback channel. (a) Top view. (b) Side view.  Varactor diodes $V_1$ and $V_2$ are identical.}
    \label{fig:Feedback}
    \end{center}
    \end{figure}

In this Article, we analyze a specific model of photo-metamaterial consisting of meta-atoms depicted schematically in Fig.~\ref{fig:Feedback}. We prove that if the intensity of the wave exciting such  meta-atom exceeds a threshold value, the meta-atom inversion symmetry is dynamically broken and as a result nonzero second-harmonic signal emerges. 

The operating principle of the proposed system is as follows. If the photodiode is not illuminated,  the meta-atom is inversion-symmetric due to back-to-back orientation of identical varactors. As a result, second-order nonlinear response of the structure vanishes. However, if the intensity of the incident wave is large enough, voltage rectified on varactor $V_2$ becomes sufficient for the light emission from light-emitting diode (LED). The photodiode illuminated by the light from LED creates static bias voltages $U_1$ and $U_2$ on varactors $V_1$ and $V_2$, and in the general case these bias voltages are different. Consequently, the inversion symmetry of the system breaks down, second-order nonlinearities from varactors $V_1$ and $V_2$ are no longer compensated and the meta-atom starts to  generate second-harmonic signal.

The rest of the paper is organized as follows. In Section~\ref{sec:Meta-atom} we analyze the  response of the individual meta-atom to the external field as a function of impinging wave intensity at fixed frequency. Section~\ref{sec:Structure} discusses the influence of the optical feedback channel on the effective susceptibilities of the entire structure and reveals their hysteresis behavior. Finally, in Sec.~\ref{sec:Discussion} we summarize the obtained results and outline the perspectives for further experimental studies. Appendix outlines the calculation of effective susceptibilities for the considered photo-metamaterial.

\section{Nonlinear response of a single meta-atom}\label{sec:Meta-atom}

First we examine the properties of a single meta-atom operating at microwave frequencies (Fig.~\ref{fig:Feedback}). The meta-atom is based on the ring with the radius $r=7.5$~mm, width  $h=1.0$~mm and thickness $b=1.0$~mm. Two identical Skyworks SMV 1405 varactors are inserted into the ring symmetrically, and the linear capacitance of the ring itself (without the capacitance of inserted varactors) is equal to $C_l=1.5$~pF. Two BPW-34S photodiodes are attached in parallel with varactor $V_1$, and L-53SRD-H Kingbright LED is inserted in parallel with varactor $V_2$ as shown in Fig.~\ref{fig:Feedback}(a). The inductance of the ring is calculated as~\cite{Grover}
\begin{equation}\label{Inductance}
L=\mu_0 r\,\left[\log\left(\frac{8\,r}{h+b}\right)-0.5\right]=27.3~\text{nH}\:.
\end{equation}
Photodiodes and LED are positioned in a close proximity. The meta-atom resonance frequency for inactive optical feedback channel is equal to $f_0=1.147$~GHz. Note that the ratio of wavelength to the meta-atom diameter at resonance frequency is $\lambda_0/(2\,r)=17$, i.e. the scatterer has subwavelength size.

Light illuminating photodiode creates irradiance $\er$ giving rise to backward current $I_{\rm{L}}(\er)$. This light-induced current charges varactor diode $V_1$ and creates a bias voltage on it. Varactor $V_1$ bias voltage $U_1$ can be found from Shockley diode equation~\refPR{Seeger}
\begin{equation}\label{BiasVolt}
I=-I_{\rm{L}}(\er)+I_{\rm{s}}\,\left[\exp\left(\frac{U_1}{n_{\rm{D}}\,N\,U_{\rm{T}}}\right)-1\right]=0\:,
\end{equation}
where ``thermal'' voltage $U_{\rm{T}}=kT/e=25.8$~mV for the temperature $T=300$~K, saturation current $I_{\rm{s}}=5.0$~nA and diode ideality factor $n_{\rm{D}}=1.4$ (for  BPW-34-S photodiode). In Eq.~\eqref{BiasVolt} varactor reverse bias is deemed positive. In order to maximize bias voltage $U_1$ we use $N=2$ photodiodes loaded in series. The dependence of light current on irradiance is approximated by the piecewise function 
\begin{equation}
\begin{split}
& I_{\rm{L}}(\er)=\eta(\lambda)\,B\,\er, \mspace{20mu} \text{for}\mspace{9 mu} \er<\er_s\:,\\
& I_{\rm{L}}(\er)=\eta(\lambda)\times 600~\mu A\,\mspace{20 mu} \text{for}\mspace{9 mu} \er\geq \er_s
\end{split}
\end{equation}
with $\er_s=12~\rm{mW}/\rm{cm}^2$, $B=50\,\mu\text{A}\cdot \text{cm}^2/\text{mW}$. $\eta(\lambda)$ is a relative spectral sensitivity of photodiode equal to 0.6 for LED light at wavelength $660$~nm. Thus,
\begin{equation}\label{BiasVolt2}
U_1=N\,n_{\rm{D}}\,U_{\rm{T}}\,\ln\left(1+\frac{I_{\rm{L}}(\er)}{I_{\rm{s}}}\right)\:.
\end{equation}
Here, the irradiance $\er$ and bias voltage $U_1$ are understood as the quantities averaged over the microwave signal period.

The nonlinearity of the meta-atom response originates from the dependence of varactor capacitance on  static bias voltage approximated by the formula
\begin{equation}\label{CU}
C(U)=\frac{C_{\rm{J}}}{(1+U/U_{\rm{J}})^M}\:,
\end{equation}
where positive $U$ describes varactor reverse bias, and $C_{\rm{J}}=2.37$~pF, $U_{\rm{J}}=0.77$~V, $M=0.5$ are empirical parameters of Skyworks SMV 1405 varactor~\cite{Skyworks}. The resistance associated with varactor is equal to $R=0.8$~Ohm. 

We rearrange Eq.~\eqref{CU} as
\begin{equation}\label{QU}
U(q)=U_{\rm{J}}\,\left[\left(1+\frac{1-M}{C_{\rm{J}}\,U_{\rm{J}}}\,q\right)^{1/(1-M)}-1\right]\:.
\end{equation}
The latter expression can be expanded in power series with in the vicinity of the point $q_1=\frac{C_{\rm{J}1}\,U_{\rm{J}1}}{M_1-1}\,\left[1-(1+U_1/U_{\rm{J}1})^{1-M_1}\right]$ corresponding to varactor $V_1$ stationary charge:
\begin{equation}\label{ExpU1}
\begin{split}
& U(\Delta q)=U_1+\left(1+U_1/U_{\rm{J}1}\right)^{M_1}\,\frac{\Delta q}{C_{\rm{J}1}}+\\
& \frac{M_1}{2\,C_{\rm{J}1}^2\,U_{\rm{J}1}}\,\left(1+U_1/U_{\rm{J}1}\right)^{2M_1-1}\,\Delta q^2+\\
& \frac{M_1\,(2M_1-1)}{6\,C_{\rm{J}1}^3\,U_{\rm{J}1}^2}\,\left(1+U_1/U_{\rm{J}1}\right)^{3M_1-2}\,\Delta q^3\:,
\end{split}
\end{equation}
where $\Delta q\equiv q-q_1$. Taking into account parasitic capacitance $C_{\rm{p}1}=0.29$~pF loaded parallel to varactor and denoting by $Q$ the total charge stored in both varactor $V_1$ and parasitic capacitance $C_{\rm{p}1}$, we obtain:
\begin{equation}\label{ExpU2}
\begin{split}
& U(\Delta Q)=U_1+\frac{\Delta Q}{C_1}+\\
& \frac{M_1\,C_{\rm{J}1}}{2\,C_1^3\,U_{\rm{J}1}}\,\left(1+U_1/U_{\rm{J}1}\right)^{-M_1-1}\,\Delta Q^2+\\
& \left[\frac{(2M_1-1)\,C_1}{6}-\frac{M_1\,C_{\rm{p}1}}{2}\right]\times \\
& \frac{M_1\,C_{\rm{J}1}}{C_1^5\,U_{\rm{J}1}^2}\,\left(1+U_1/U_{\rm{J}1}\right)^{-M_1-2}\,\Delta Q^3\:,
\end{split}
\end{equation}
where $C_1=C_{p1}+C_{J1}\,(1+U_1/U_{J1})^{-M_1}$ is the total linear capacitance of biased varactor plus parasitic capacitance. Note that in the absence of excitation varactor $V_2$ is also biased and its bias voltage $U_2$ can be found from the equation
\begin{equation}\label{BiasVolt3}
\begin{split}
& \frac{1}{C_{\rm{l}}}\,\left\lbrace C_{\rm{p}2}\,U_2+\frac{C_{\rm{J}2}\,U_{\rm{J2}}}{M_2-1}\,\left[1-\left(1+\frac{U_2}{U_{\rm{J}2}}\right)^{1-M_2}\right]\right\rbrace=\\
& U_1-U_2
\end{split}
\end{equation}
where $C_l$ is the linear capacitance associated with SRR meta-atom and subscript index $2$ denotes the quantities that refer to the second varactor $V_2$. It should be emphasized that it is finite linear capacitance $C_{\rm{l}}$ that gives rise to the difference between $U_1$ and $U_2$ leading eventually to the nonzero second-order nonlinear susceptibility.

Current-voltage characteristics of L-53SRD-H Kingbright LED is approximated by the formula
\begin{equation}\label{LED}
I_{\rm{LED}}=\frac{\bar{U}_{\rm{LED}}-U_{\rm{t}}}{R_{\rm{LED}}}
\end{equation}
where $R_{\rm{LED}}=15.6$~Ohm is LED effective resistance, $U_{\rm{t}}=1.9$~V is LED threshold voltage and $\bar{U}_{\rm{LED}}$ is the voltage on LED averaged over the period of microwave signal. The luminous intensity produced by LED is directly proportional to forward current
\begin{equation}\label{LuminousI}
\il=K\,I_{\rm{LED}}
\end{equation}
where $K=1.56\,\text{W}/(\text{sr}\cdot\text{A})$, and the wavelength of light emitted by LED is equal to $660$~nm.

    \begin{figure}[b]
    \begin{center}
    \includegraphics[width=0.8\linewidth]{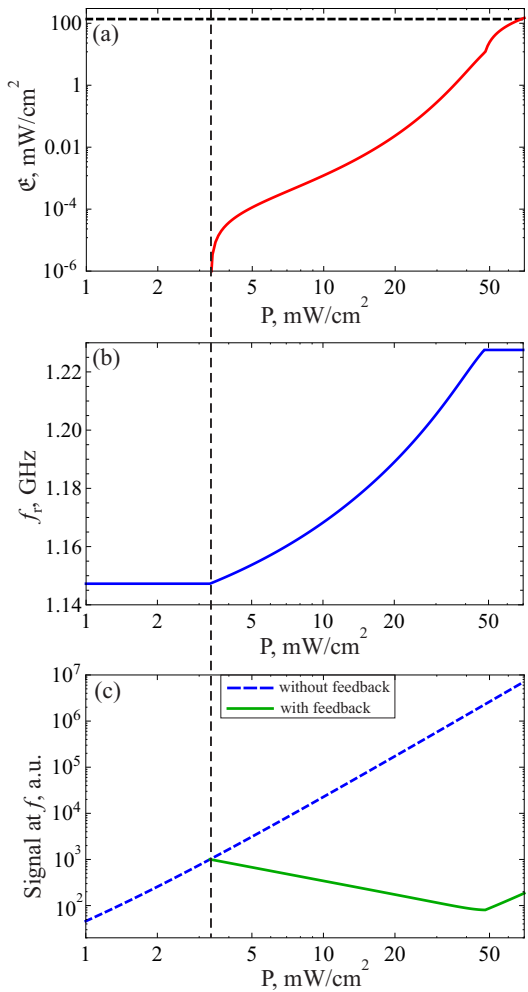}
    \caption{Response of the individual meta-atom to the external excitation at fixed frequency $f=1.12$~GHz. (a) Irradiance of the photodiode. Horizontal dashed line indicates the irradiance created by the bright sunlight; (b) resonance frequency of the individual meta-atom; (c) intensity of the scattered field at frequency $f$ as a function of impinging wave intensity. Vertical dashed line indicates the threshold intensity at given frequency.}
    \label{fig:Irradiance}
    \end{center}
    \end{figure}

To find the resulting irradiance of photodiodes, we assume that all light emitted by LED illuminates the photo-sensitive surface of photodiodes. This gives an equation
\begin{equation}\label{SelfConsEq}
2\pi\,\left[1-\cos\theta\right]\,\il=N\,w\,l\,\er\:,
\end{equation}
where $\il$ is LED luminous intensity, $2\,\theta=60^{\circ}$ is viewing angle for LED,  $w\times l$ is the size of a single photodiode, $N$ is the number of photodiodes loaded in series and $\er$ is the average photodiode irradiance.

With this model, we analyze the response of the meta-atom to the incident field with the frequency $f=1.120$~GHz slightly below the meta-atom resonance and with intensity varying from zero to $70\,\text{mW}/\text{cm}^2$. At intensities lower than the threshold value $P_c=3.33\,\text{mW}/\text{cm}^2$ the optical feedback channel is inactive since the voltage rectified on varactor $V_2$ is insufficient to set light to LED. Therefore, meta-atom operates as a typical passive nonlinear element with zero second-order nonlinear susceptibility (Fig.~\ref{fig:Irradiance}, see the fragment before the vertical dashed line).

    \begin{figure}[b]
    \begin{center}
    \includegraphics[width=0.8\linewidth]{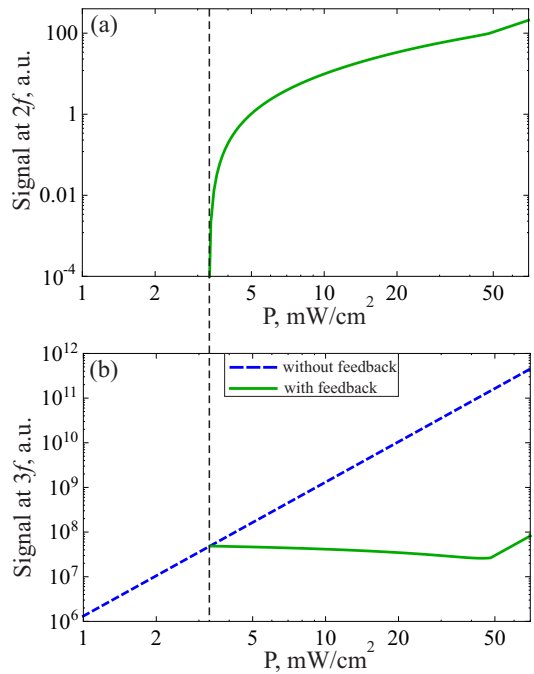}
    \caption{Intensity of harmonics scattered by the meta-atom as a function of impinging wave intensity. (a) Second harmonic signal; (b) third harmonic signal. Frequency of impinging wave is equal to $f=1.12$~GHz. Vertical dashed line indicates the threshold intensity at given frequency.}
    \label{fig:Harmonics}
    \end{center}
    \end{figure}

When the intensity of excitation exceeds the threshold value $P_c$, the luminous intensity produced by LED (as well as photodiode irradiance) starts to grow almost exponentially as a function of $(P-P_c)$ [Fig.~\ref{fig:Irradiance}(a)]. This is accompanied by the linear growth of the meta-atom resonance frequency [Fig.~\ref{fig:Irradiance}(b)]. The upward resonance shift is explained by the decrease in the capacitance of varactors due to the reverse bias created by the photodiodes.  Note that if the polarity of photodiodes is switched, the resonance frequency will experience a decrease. At the same time, optical feedback channel suppresses the scattering at the fundamental frequency $f$ as illustrated in Fig.~\ref{fig:Irradiance}(c). The obtained result is well understandable: since the resonance shifts away from the fixed frequency of excitation, the effective meta-atom polarizability decreases thus suppressing scattering even though the intensity of excitation increases [see the solid curve in Fig.~\ref{fig:Irradiance}(c)].

Finally, after reaching some irradiance $\er_s$ the response of the photodiode saturates, i.e. bias voltage produced by the photodiode no longer increases with irradiance. The meta-atom resonance frequency then stabilizes at $f_{\rm{max}}=1.228$~GHz, which is $81$~MHz higher than the resonance frequency without optical feedback. Accordingly, the intensity of the scattered field at this stage grows with the intensity of excitation.

At the same time, the meta-atom generates harmonics with the frequencies $2\,f$, $3\,f$, etc. as shown in Fig.~\ref{fig:Harmonics}. Quite remarkably, the dependence of harmonic signal intensity on intensity of impinging wave is not captured by the simple power law formula in contrast with the  ``standard'' nonlinear structures. Such deviations from the power law should be attributed to the presence of feedback channel that gives rise to second-harmonic generation for $P>P_c$ [Fig.~\ref{fig:Harmonics}(a)] and simultaneously suppresses third-harmonic generation [Fig.~\ref{fig:Harmonics}(b)].

\section{Hysteresis behavior and nonlinearities of the composite structure}\label{sec:Structure}

\begin{figure}[b]
    \begin{center}
    \includegraphics[width=1.0\linewidth]{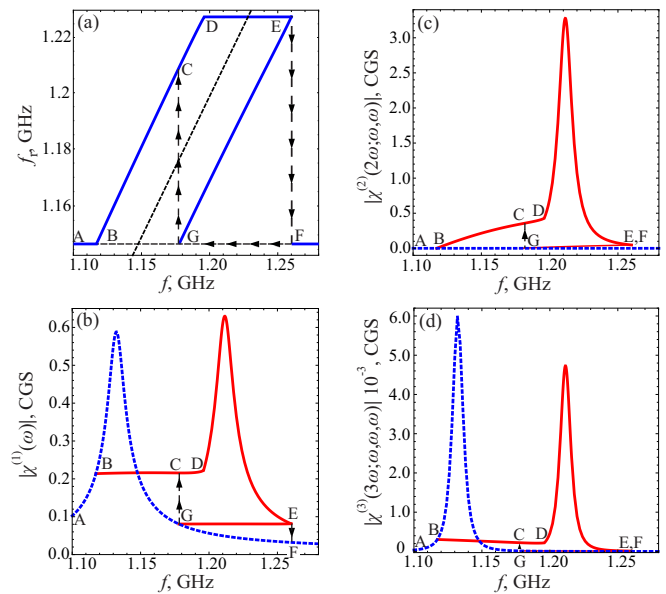}
    \caption{(a) Dependence of the meta-atom resonance frequency on frequency of the incident wave  when the intensity of excitation is fixed and equal to $P=4.0\,\text{mW}/\text{cm}^2$. Dashed line corresponds to the resonance condition $f_r=f$. (b-d) Effective linear and nonlinear susceptibilities of the metamaterial based on the discussed nonlinear meta-atoms with optical feedback channel. A hysteresis behavior is observed. (b) Linear susceptibility. (c,d) Second- and third-order nonlinear susceptibilities.}
    \label{fig:Chi13}
    \end{center}
    \end{figure}

To grasp the physics governing the behavior of the composite structure, it is instructive to plot the dependence of the single meta-atom resonance frequency on frequency of excitation for a fixed intensity of the incident wave $P=4.0\,\text{mW}/\text{cm}^2$ [Fig.~\ref{fig:Chi13}(a)].

First we describe the behavior of the meta-atom when the frequency of excitation is gradually increased. For the frequencies below $f<1.117$~GHz optical feedback channel is not active since the intensity of excitation is too low [point A in Fig.~\ref{fig:Chi13}(a)]. At the point B of the diagram LED starts working and the resonance frequency increases with the increase of frequency. However, despite the growth of LED luminous intensity the system remains below the resonance following the BD segment of the diagram Fig.~\ref{fig:Chi13}(a). When the photodiode saturation is reached ($\er=\er_s$), the resonance frequency no longer changes and the meta-atom finally reaches the resonance (DE segment). Next system reaches point E of the diagram ($f_E=1.260$~GHz) becoming  off-resonant again, but the frequency of excitation is now {\it above} the resonance frequency. At this point, the meta-atom appears to be so far from the resonance, that the intensity of excitation becomes insufficient to feed the LED, and optical feedback channel stops working. The meta-atom falls into the state F. Thus, the overall route of the system is represented by the path ABCDEF in the diagram.

The meta-atom behavior becomes quite different when excitation frequency is gradually decreased. At frequencies above $f_F=1.260$~GHz LED does not work. With the decrease of the driving field frequency the system moves to the point G of the diagram ($f_G=1.178$~GHz). At this point the intensity of excitation reaches the threshold value $P_c$ and LED starts working. The meta-atom thus  ``jumps'' into the state $C$. Further decrease of the excitation frequency brings the system into B point, and at this moment the feedback channel switches off. The overall route of the system in this scenario is represented by the path FGCBA in the diagram Fig.~\ref{fig:Chi13}(a).

Thus, the system exhibits a bistable behavior. Note that the meta-atom actually ``skips'' the resonance in the scenario when the driving frequency is gradually decreased, whereas the system passes through resonance in the opposite case of frequency increase.

This simple reasoning suggests that the effective susceptibilities of the entire photo-metamaterial exhibit a bistable behavior. To confirm this idea, we calculated the effective susceptibilities of the metamaterial with rectangular lattice $3.0\times 3.0\times 2.0$~cm ($2.0$~cm is the period along the axis of the ring) as specified in Appendix. The obtained results are presented in Fig.~\ref{fig:Chi13}(b-d). In the case when the frequency of excitation is gradually increased, the system follows the path ABCDEF in those diagrams, whereas in the opposite scenario of frequency decrease the system follows the trajectory FGCBA without passing the resonance peak.

It should be stressed that the effective susceptibilities discussed above depend implicitly on the intensity of excitation $P$ in contrast with the ``standard'' nonlinear structures. Indeed, the effective susceptibilities depend on the meta-atom resonance frequency which depends on photodiode irradiance $\er$. The latter in turn depends on the intensity of excitation. Actually, it is the implicit dependence of susceptibilities on intensity of excitation that leads to the non-power-law dependence of generated second and third harmonics on the driving field [see Sec.~\ref{sec:Meta-atom}].

On the other hand, the perturbative treatment of nonlinearities is still possible since the nonlinear polarization at frequencies $2\,f$ or $3\,f$ is around $10\%$ of the polarization at the fundamental frequency $f$. The technical details of effective susceptibilities evaluation are specified in Appendix.

\section{Discussion and outlook}\label{sec:Discussion}

In conclusion, we have demonstrated a rich physics originating from the combination of such concepts as optical feedback and dynamic symmetry breaking. The meta-atoms of the proposed structure can dynamically acquire second-order nonlinear response due to the dynamic inversion symmetry breaking. We prove that the dependence of harmonic signals on intensity of excitation is not captured by the standard power law formula, and the entire structure features the hysteresis behavior of effective susceptibilities.

We expect that such photo-metamaterials may exhibit an intriguing physics due to optical feedback. For instance, in the two-pulse interaction regime the first pulse (pump) may dynamically turn on and off second-order nonlinearity while passing through the metamaterial, whereas the second pulse (probe) will experience the modulated nonlinear properties in the course of propagation. Even more exotic effects are expected if the optical feedback couples photodiodes and LEDs of the {\it different} meta-atoms which is a promising direction for further studies.

Our calculations for the realistic microwave metamaterials demonstrate that the predicted effects including dynamic inversion symmetry breaking can be realized and investigated in the proof-of-concept microwave experiments. A similar physics can be also observed in other spectral ranges if the appropriate mechanism for the feedback channel is implemented.

\acknowledgments 

We acknowledge valuable discussions with P. Kapitanova, G. Solomakha and Y. Kivshar. This work was supported by the Ministry of Education and Science of the Russian Federation (Zadanie No. 3.2465.2017/4.6 and 3.8891.2017/8.9) and the Australian Research Council (DP~150103611).

\begin{widetext}
\section*{Appendix}

To analyze the arising collective response of photo-metamaterial, we derive effective linear and nonlinear susceptibilities. We describe the properties of the individual meta-atom in terms of its linear and nonlinear polarizabilities $\alpha_1$, $\alpha_2$ and $\alpha_3$ as follows:
\begin{gather}
m(\om)=\alpha_1(\om)\,H(\om)+2\,\alpha_2(\om;2\,\om,-\om)\,H(2\om)\,H^*(\om)+3\,\alpha_3(\om;\om,\om,-\om)\,H^3(\om)\:,\label{Mag1}\\
m(2\,\om)=\alpha_1(2\,\om)\,H(2\,\om)+\alpha_2(2\,\om;\om,\om)\,H^2(\om)\:,\label{Mag2}\\
m(3\,\om)=\alpha_1(3\,\om)\,H(3\,\om)+2\,\alpha_2(3\,\om;2\,\om,\om)\,H(2\,\om)\,H(\om)+\alpha_3(3\,\om;\om,\om,\om)\,H^3(\om)\:,\label{Mag3}
\end{gather}
where $m(\om)$ is the meta-atom magnetic moment and $H(\om)$ is the magnetic field acting on the particle. The terms proportional to $H^4(\om)$, $H(3\,\om)\,H(\om)$, $H^2(2\,\om)$ and higher-order terms are omitted. The meta-atom polarizabilities $\alpha_1$, $\alpha_2$ and $\alpha_3$ depend on the irradiance of photodiode $\er$.

We compute the effective polarizabilities for the discussed meta-atom (Fig.~\ref{fig:Feedback}) below. In this calculation we assume that the current flowing through LED is negligible if compared with the displacement current in varactor. Under this assumption we derive  nonlinear oscillator equation for the charge accumulated in varactors:
\begin{equation}\label{NonlinOsc}
\Delta\ddot{Q}+2\,\beta_0\,\Delta\dot{Q}+\omega_0^2\,\Delta\,Q+\beta_2\,\Delta\,Q^2+\beta_3\,\Delta\,Q^3=-\frac{S}{c}\,\dot{B}(t)
\end{equation}
where $S=\pi\,r^2$ is the area limited by the ring, $\beta_0=R/(2\,L)$, $R$ is the net resistance of meta-atom, $\omega_0^2=1/(LC)$, $C^{-1}=C_1^{-1}+C_2^{-1}+C_l^{-1}$, $\beta_2=a_1-a_2$, $\beta_3=b_1+b_2$ with 
\begin{equation}
a_{1,2}=\frac{M_{1,2}\,C_{\rm{J}1,2}}{2L\,C_{1,2}^3\,U_{\rm{J}1,2}}\,\left(1+\frac{U_{1,2}}{U_{\rm{J}1,2}}\right)^{-M_{1,2}-1}\:,
\end{equation}
\begin{equation}
b_{1,2}=\left[\frac{(2M_{1,2}-1)\,C_{1,2}}{6}-\frac{M_{1,2}\,C_{\rm{p}1,2}}{2}\right]\,\frac{M_{1,2}\,C_{\rm{J}1,2}}{C_{1,2}^5\,U_{\rm{J}1,2}^2\,L}\,\left(1+\frac{U_{1,2}}{U_{\rm{J}1,2}}\right)^{-M_{1,2}-2}\:.
\end{equation}
Eq.~\eqref{NonlinOsc} can be solved by means of standard perturbation theory~\refPR{Boyd} searching the steady-state  solution as a series
\begin{equation}\label{SeriesPerturb}
\Delta\,Q=\zeta\,Q^{(1)}(t)+\zeta^2\,Q^{(2)}(t)+\zeta^3\,Q^{(3)}(t)\:,
\end{equation}
where $\zeta$ parameter is related to the amplitude of impinging field $H(\om)$. Putting the anzatz Eq.~\eqref{SeriesPerturb} into Eq.~\eqref{NonlinOsc} and separating the equations for the different powers of $\zeta$, we obtain a set of linear differential equations with unknown functions $Q^{(1)}$, $Q^{(2)}$ and $Q^{(3)}$, respectively. We further evaluate the magnetic moment of the meta-atom as $m=\Delta\dot{Q}\,S/c$ and extract the relevant polarizabilities:
\begin{gather}
\alpha_1(\omega)=\frac{\omega^2\,S^2}{c^2\,L\,D(\omega)}\:,\label{Al1}\\
\alpha_2(2\omega;\omega,\omega)=-\frac{i\omega^3}{c^3}\,\frac{\beta_2\,S^3}{L^2\,D^2(\omega)\,D(2\omega)}\:,\label{Al2}\\
\alpha_3(3\omega;\omega,\omega,\omega)=\frac{3\,\omega^4\,S^4}{c^4\,L^3}\,\left(\frac{\beta_2^2}{2\,D^3(\omega)\,D(2\omega)\,D(3\,\omega)}-\frac{\beta_3}{4\,D^3(\omega)\,D(3\,\omega)}\right)\:.\label{Al33}\\
\alpha_3(\om;\om,\om,-\om)=\frac{\om^4\,S^4}{c^4\,L^3\,|D(\om)|^2\,D^2(\om)}\,\left(-\frac{3\,\beta_3}{4}+\frac{\beta_2^2}{\om_0^2}+\frac{\beta_2^2}{2\,D(2\,\om)}\right)\:.\label{Al31}
\end{gather}
where $D(\omega)=\omega_0^2-2i\,\beta_0\,\omega-\omega^2$, $S=\pi\,r^2$ and $L$ is a total inductance of the circuit.

Solving the equation for irradiance \eqref{SelfConsEq} numerically, we calculate the values of effective polarizabilities Eqs.~\eqref{Al1}-\eqref{Al31} and estimate effective susceptibilities of the composite structure as follows~\cite{Boyd}:
\begin{gather}
\chi^{(1)}_{\rm{loc}}(\omega)=\frac{\alpha_1(\omega)/a^3}{1-4\,\pi\,\alpha_1(\omega)/(3\,V_0)}\:,\label{Chi1Loc}\\
\chi^{(2)}_{\rm{loc}}(2\,\omega;\omega,\omega)=\frac{\alpha_2(2\,\omega;\omega,\omega)}{V_0}\,\frac{\varepsilon_{\rm{loc}}(2\,\omega)+2}{3}\,\left[\frac{\varepsilon_{\rm{loc}}(\omega)+2}{3}\right]^2\:,\label{Chi2Loc}\\
\chi^{(3)}_{\rm{loc}}(3\,\omega;\omega,\omega,\omega)=\frac{\alpha_3(3\,\omega;\omega,\omega,\omega)}{V_0}\,\frac{\varepsilon_{\rm{loc}}(3\,\omega)+2}{3}\,\left[\frac{\varepsilon_{\rm{loc}}(\omega)+2}{3}\right]^3\:,\label{Chi3Loc}
\end{gather}
where $V_0$ is the unit cell volume ($18\,\text{cm}^3$ in the calculations above) and $\varepsilon_{\rm{loc}}(\om)=1+4\,\pi\,\chi^{(1)}_{\rm{loc}}(\om)$. The described procedure yields effective susceptibilities which depend on irradiance $\er$. Note that the simplified formulas Eqs.~\eqref{Chi1Loc}-\eqref{Chi3Loc} completely ignore spatial dispersion, and they are used here for the estimations only. A more complete approach incorporating spatial dispersion and based on the discrete dipole model is presented in Ref.~\cite{Voytova}.

\end{widetext}

\bibliography{NonlinearLib}

\end{document}